# Reactionaries and Einstein's Fame: "German Scientists for the Preservation of Pure Science," Relativity, and the Bad Nauheim Meeting

Jeroen van Dongen

*Einstein Papers Project*
*California Institute of Technology*
*Pasadena CA 91125, USA*

*Institute for History and Foundations of Science*
*Utrecht University*
*P.O. Box 80.000*
*3508 TA Utrecht, the Netherlands*

Two important and unpleasant events occurred in Albert Einstein's life in 1920: That August an antirelativity rally was held in the large auditorium of the Berlin Philharmonic, and a few weeks later Einstein was drawn into a tense and highly publicized debate with Philipp Lenard on the merits of relativity at a meeting in Bad Nauheim, Germany. I review these events and discuss how they affected Einstein in light of new documentary evidence that has become available through the publication of Volume 10 of the *Collected Papers of Albert Einstein*.

## Introduction

On August 27, 1920, the front page of the *Berliner Tageblatt*, a widely-read Berlin daily newspaper, carried a statement by Albert Einstein (1879-1955):

> Under the pretentious name "Arbeitsgemeinschaft deutscher Naturforscher" ["Working Society of German Scientists"], a variegated society has assembled whose provisional purpose of existence seems to be to degrade, in the eyes of nonscientists, the theory of relativity, as well as me as its originator.... I have good reason to believe that motives other than the striving for truth are at the bottom of this business. (If I were a German nationalist with or without a swastika instead of a Jew with liberal international views, then...).[1]

Just one year earlier, Einstein had been celebrated as the Newton of his age for his greatest scientific triumph, the confirmation of his prediction, based upon his general theory of relativity, that light would be bent by the gravitational field of the sun. This observation by the British solar eclipse expedition under Arthur S. Eddington (1882-1944) had caught the imagination of the public through extensive coverage by newspapers around the world, instantly propelling Einstein to international fame.[2] The huge public acclaim and adulation that was accorded Einstein--a democrat, pacifist and Jew--upset the reactionary circles of the Weimar Republic. It also vexed conservative academics, notably the Nobel Laureate Philipp Lenard (1862-1947), who may have felt that the theoretical physicist Einstein had captured too much of the limelight, while other, experimental physicists such as himself were not appreciated enough.[3]

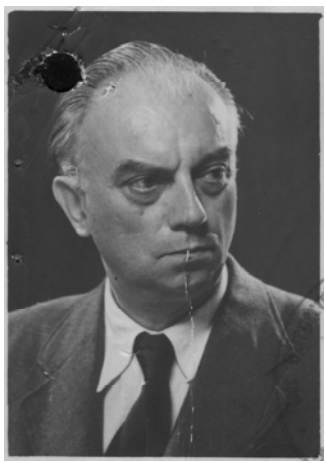

Figure 1: Paul Weyland (1888-1972). *Source*: Andreas Kleinert, Halle, private copy.

Einstein experienced the harsh consequences of his newly acquired world-wide fame in the summer of 1920: In August, in the large auditorium of the Berlin Philharmonic (which had a capacity of over 1600 people), a series of lectures was staged that denounced him as a fraud and a propagandist, and in September, at a meeting of the Society of German Scientists and Physicians (*Gesellschaft Deutscher Naturforscher und Ärzte*) in Bad Nauheim, Germany, he became embroiled in an intense debate with Lenard on the merits of his theory of relativity.[4] I will review these events, emphasizing Einstein's reactions and emotions as revealed in his correspondence of the period through the publication of Volume 10 of the *Collected Papers of Albert Einstein*.[5]

**The Anti-Einstein Rally at the Berlin Philharmonic**

The anti-Einstein campaign kicked off on August 6, 1920, with an inflammatory article in the *Tägliche Rundschau*, a Berlin daily newspaper: "Herr Albertus Magnus has been resurrected";[6] he has stolen the work of others and has mathematized physics to such an extent that fellow physicists have been left clueless. Furthermore, the article continued, Einstein had undertaken a propaganda campaign by which he had cast a spell both over the public and over academic circles--but in reality relativity was nothing but fraud and fantasy. The author of the piece was Paul Weyland (1888-1972, figure 1), an obscure right-wing publicist and talented rabble-rouser--one of the shadier products of postwar Berlin.[7]

Most of Weyland's accusations were not new. His charges of plagiarism and propaganda had been leveled earlier by Ernst Gehrcke (1878-1960), a prominent spectroscopist at the prestigious Imperial Physical-Technical Institute (*Physikalisch-Technische Reichsanstalt*) in Berlin-Charlottenburg.[8] Weyland also drew heavily on Lenard's more substantive objections to Einstein's theory of relativity, which Lenard had published in 1918.[9] Einstein had responded immediately to Lenard's criticisms,[10] but Weyland contended that they had remained undisputed.

Weyland's shrill tone in his newspaper article and the highly public character of his accusations were indeed new, however. Also new was their thinly concealed anti-Semitic character: Weyland claimed that Einstein had "a particular press, a particular community [*Gemeinde*]" that kept feeding pro-Einstein stories to the public.[11] A word to the wise was enough: The widely circulating, liberal *Berliner Tageblatt* was published by Rudolph Mosse (1843-1932), a prominent Jew, and was known in anti-Semitic circles as the *Judenblatt*--the Jew paper.[12] In 1919 it had carried an article announcing the results of the British solar eclipse expedition that rose to laudatory hyperbole, not shying away from declaring that "a highest truth, beyond Galileo and Newton, beyond Kant" had been unveiled by "an oracular saying from the depth of the skies."[13] Its author, Alexander Moszkowski (1851-1934), was a close acquaintance of Einstein's--and also Jewish.[14] Further, Einstein himself had published a short note on the results of the British eclipse expedition in *Die Naturwissenschaften*,[15] a highly visible journal whose editor-in-chief was Arnold Berliner (1862-1942)--another Jew.[16] Finally, on December 14, 1919, the front page of the *Berliner Illustrirte Zeitung* carried a large close-up portrait of Einstein (figure 2) whose caption read: "A new eminence in the history of the world: Albert Einstein, whose researches signify a complete revolution of our understanding of Nature and whose insights equal in importance those of a Copernicus, Kepler, and Newton."[17] This newspaper had been founded by Leopold Ullstein (1826-1899), yet another prominent Berlin Jew.[18] Weyland thus had seen enough of Einstein's methods: If "German science" now decided to close ranks and take action against Einstein, "settling scores," Einstein had only himself to blame.[19]

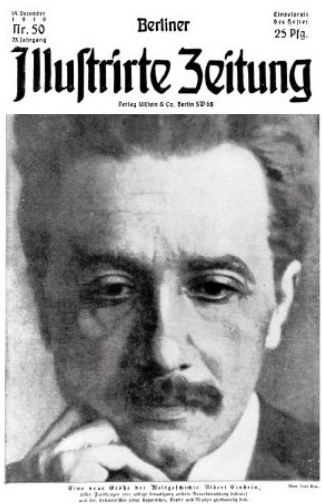

Figure 2: Albert Einstein on the first page of the Berliner Illustrirte Zeitung of December 14, 1919.

On August 11, 1920, five days after Weyland's diatribe had appeared in the *Tägliche Rundschau*, Max von Laue (1879-1960), another Nobel Laureate and Einstein's colleague and friend in Berlin, replied to it in the same newspaper.[20] Von Laue denied categorically that Einstein had plagiarized the work of others and made quite clear that Einstein fully deserved the praise accorded to him by his fellow physicists. Von Laue also aired his own objections to Lenard's criticisms of the theory of relativity that Weyland had quoted--to which Weyland responded immediately.[21] At the same time, Weyland announced that a series of lectures would be held that would reveal the true Einstein. Indeed, notices soon appeared in *Der Tag* and in the *Tägliche Rundschau* announcing twenty lectures, organized by the Working Society of German Scientists for the Preservation of Pure Science (*Arbeitsgemeinschaft deutscher Naturforscher zur*

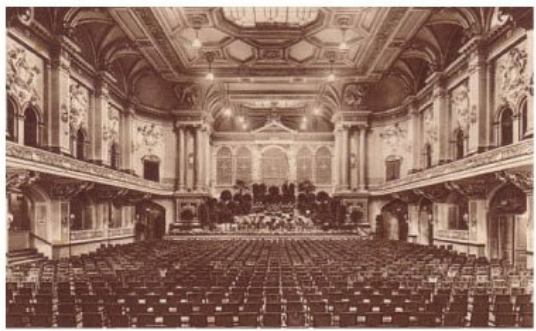

Figure 3: The large auditorium of the Berlin Philharmonic as seen in a picture of 1939.

*Erhaltung reiner Wissenschaft e.V.*[*]) to be held in the large auditorium of the Berlin Philharmonic beginning on August 24, with Weyland and Gehrcke as the first speakers.[22] This Working Society in all likelihood comprised only Weyland himself,[23] who clearly had invented it as a front to provide a veneer of respectability and to suggest that his efforts had wide support among scientists.

Such tactics were not too surprising in light of Weyland's political credentials: He was an active member of the German National People's Party (*Deutschnationale Volkspartei*, DNVP), a right-wing political party whose membership ranged from nationalistic conservatives to *völkisch* racists. Later, in October 1920, Weyland also published an article in the right-wing *Deutsche Zeitung*, speaking out against the DNVP's position on the "Jewish question"--which was much too soft for his taste. Indeed, physicist-historian Hubert Goenner believes that Weyland undertook his anti-Einstein campaign to garner support for the various far right-wing groups with which he was involved.[24] In any case, Einstein was impressed enough with Weyland's diatribes to believe that he indeed had the endorsement of a fair number of physicists.[25]

As Weyland had announced, on August 24, 1920, he delivered a demagogical rant in the large auditorium of the Berlin Philharmonic (figure 3):

> *Meine Damen und Herren!* Hardly ever in science has a scientific system been set up with such a display of propaganda as the general principle of relativity, which on closer inspection turns out to be in the greatest need of proof.[26]

He deplored that propaganda, giving a cascade of citations from the *Einsteinpress*. He castigated especially the coverage of the results of the British eclipse expedition in *Die Naturwissenschaften*: "from this headquarters [of the Einstein press] popular sentiment is roused for him."[27] He heaped scorn on comparisons with "Copernicus, Kepler, Newton," such as those in the *Berliner Tageblatt* and *Berliner Illustrirte Zeitung*. Einstein himself was to blame for this: "By a single word" to this press, "which has excellent connections with his circles," he could put an end to this "wave of glorification and admiration."[28] Weyland cast his net even wider than the Jewish-owned or edited newspapers: He railed against a number of science popularizers (who may or may not have been Jewish), and he misquoted work of the philosopher Moritz Schlick (1882-1936).[29] All of which proved that Einstein's theory of relativity was simply fad and fiction.

---

[*] The abbreviation *e.V.* for *eingetragener Verein* purports to claim that it is an officially registered society.

Weyland maintained that Einstein had resorted to propagandistic methods when academic criticism had risen too high--and at the same time he had obstructed and suppressed academic critics. All of this was a consequence of the intellectual and moral decay of German society, which was being exploited and promoted by "a certain press." This decay had "attracted all kinds of adventurers, not only in politics, but also in art and science." Indeed, the theory of relativity was being "thrown to the masses" in exactly the same way as "the dadaist gentlemen" promoted their wares, which had to do as little as relativity with the observation of nature. No one should be surprised, therefore, Weyland concluded, that a movement had arisen to counter this "scientific Dadaism" [figure 4].[30]

Einstein was present at Weyland's lecture and seemed to be unshaken by it,[31] but it must have upset him. Ernst Gehrcke spoke after Weyland: He too claimed that relativity was nothing but "scientific mass hypnosis"; it was inconsistent, led to solipsism, and was not confirmed by observation--but at least Gehrcke did not rant and rave.[32] To Einstein, it was completely clear that his opponents were politically motivated.[33] According to one account, anti-Semitic pamphlets were handed out, and according to another, swastika lapel pins were being sold.[34] Lenard's antirelativistic booklet was also being sold. Not surprisingly, Lenard was scheduled as one of the twenty future lecturers in Weyland's series.[35] Weyland had visited Lenard at the University of Heidelberg in early August, and Lenard's correspondence reveals that he was quite taken in by this fiery young man who "wants to combat [Einstein] systematically--as un-German."[36] In the end, however, Lenard decided against becoming too closely associated with Weyland and did not give a lecture at the Berlin Philharmonic.

The assault on Einstein was hardly an isolated incident but part of a much broader pattern in Germany at this time. The new Weimar Republic was in a volatile position: In March 1920, the reactionary Kapp *Putsch* had tried to overthrow the government of social democrat Friedrich Ebert (1871-1925), which after failing to do so generated a wave of radical right-wing violence in the summer of 1920 that appeared to focus not just on left-wing revolutionaries, but more broadly on pacifists and others who were deemed to be "traitors" to Germany. Thus, veiled assassination threats were made against the physician and pacifist Georg Friedrich Nicolai (1874-1964) in the right-wing press.[37] Einstein had spoken out on Nicolai's behalf earlier in 1920, and Nicolai's lectures at the University of Berlin, as well as Einstein's, had been disrupted by right-wing extremists.[38]

The anti-Einstein rally at the Berlin Philharmonic was widely covered, sometimes provocatively, in the Berlin press.[39] On August 27, Einstein responded indignantly to Weyland's campaign in the aforementioned article in the *Berliner Tageblatt*, directing his sharp pen not only against Weyland, but also against Gehrcke and Lenard.[40] The word circulated immediately that Einstein wanted to leave Berlin and Germany.[41]

That same day letters expressing outrage and in support of Einstein began pouring in to him from all realms of society--from Toni Schrodt (who described herself as "just a middle-class wage-earning girl") to Countess Elsa von Schweinitz und Krain;[42] from Jews and a pastor to students and professors. A certain Ina Dickmann, who had been present at the lectures at the Berlin Philharmonic, wrote that she was deeply ashamed that I had to experience that the method of political elections is now also used to club to death undesirable thinkers. I have never seen German representatives of science sink so low as on that evening.[43]

Dickmann's letter continued by attesting to the infatuation of the general public with the theory of relativity and its creator (which, as we have seen, was one of Weyland's bitterest complaints):

> Relativity theory has opened to me a philosophical world of infinite breadth. Before my inner eyes, the barriers and borders of philosophical systems came tumbling down and for the time being my spirit floats, without seeing a horizon, in the new world.... Perhaps there are many other laymen who have experienced the same, who like me feel pain and anger that in an already much-taunted Germany, a German greatness has been trampled in the dirt before the German people.[44]

Note that Dickmann appealed to Einstein as a *German* scientist. Indeed, many of the letter writers pleaded with Einstein not to leave Germany, particularly since the country was now in such troubled times.[45]

Einstein considered leaving Germany largely because he believed that Weyland had wide support for his views in scientific circles.[46] To the contrary, however, many of Einstein's colleagues rallied around him. The Berlin newspapers already had carried a statement in his support by his colleagues and friends von Laue, Heinrich Rubens (1865-1922) and Walther Nernst (1864-1941).[47] Others such as Max Planck (1858-1947) and Fritz Haber (1868-1934) wrote to him personally and urged him not to leave Berlin.[48] Paul Ehrenfest (1880-1933), Einstein's close friend and professor in Leiden, assured Einstein that if he did decide to leave Germany, arrangements likely could be made to secure a full professorship for him in The Netherlands.[49] Ehrenfest also told Einstein in no uncertain terms that his reply to his critics in the *Berliner Tageblatt* had been a mistake:

> This piece, *Meine Antwort*, is written such that I and my wife could not believe that you had written that *with your own hands*.... It contains all kinds of un-Einsteinian details that ... show that these damn pigs must have succeeded in hurting you deeply.[50]

Arnold Sommerfeld (1868-1971), President of the German Physical Society (*Deutsche Physikalische Gesellschaft*, DPG), also wrote to Einstein, imploring him not to "flee the flag,"[51] and informing him of his plans for the forthcoming meeting of the Society of German Scientists and Physicians in Bad Nauheim: He intended to ask its chairman, the Munich internist Friedrich von Müller (1858-1941), to speak out against demagoguery in his opening address. Further, he was about to mediate between Einstein and Lenard--which was imperative, given the tensions in the DPG between its members in Berlin and those not living in the German capital--in particular Johannes Stark (1874-1957), Wilhelm Wien (1864-1928), and of course Lenard. Sommerfeld anticipated that these two groups would face off at the meeting in Bad Nauheim, since possible reforms of the DPG would be discussed there.[52]

Stark and Lenard stood politically on the far right and did not support the Weimar Republic; Wien too was a conservative nationalist.[53] Einstein, by contrast, felt that "his political wishes had come true,"[54] and enjoyed something of a personal rapport with the Weimar

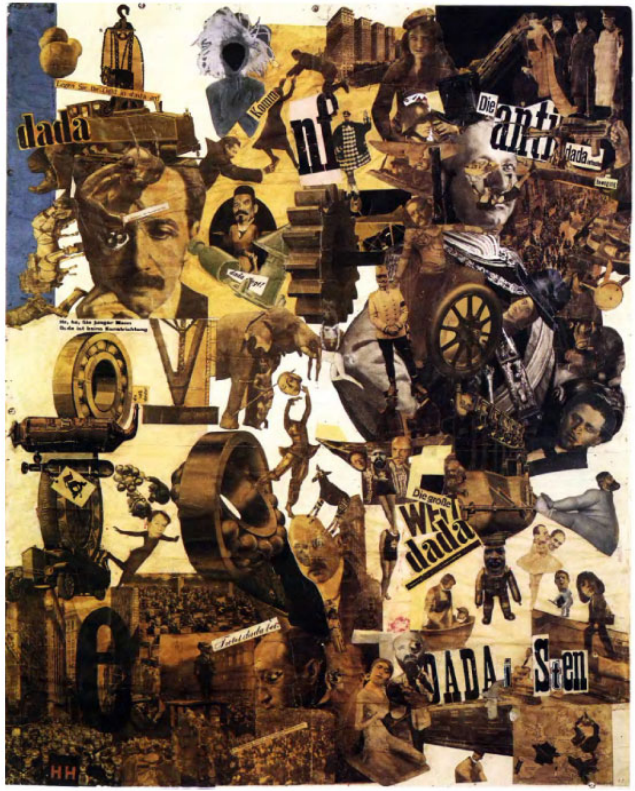

Figure 4: In this collage, which documents the hyped Berlin newspaper culture of the Weimar years, Hannah Höch used the picture of Einstein on the front cover of the *Berlinier Illistrirte Zeitung* of December 14, 1919. The collage, which is entitled *Schnitt mit dem Küchenmesser, Dada durch die letzte Weimarer Bierbauchkulturepoche Deutschlands* (*Cut with a kitchen knife, Dada through the last Weimar beer belly culture of Germany*, 1919/1920), was featured in a Dada exhibition from June 30 to August 25 in art dealer's Otto Burchard's gallery in the center of Berlin. Many press reports expressed disgust, and the exhibition was even subjected to a lawsuit for defamation of the German army; see Ludger Derenthal, "Dada, die Toten und die Überlebenden des Ersten Weltkriegs", *Zeitenblicke* 3, No. 1 (June 9, 2004). The collage not only featured Albert Einstein, but many other public figures, including Weimar's first president, Friedrich Ebert, the reactionary Putschist Wolfgang Kapp, Kaiser Wilhelm II, socialists, Dadaists, Marx and Lenin; see Gertrud Julia Dech, *Schnitt mit dem Küchenmesser, Dada durch die letzte Weimarer Bierbauchkulturepoche Deutschlands: Untersuchung zur Fotomontage bei Hannah Höch* (Münster: Lit Verlag, 1981). We do not know if Paul Weyland visited the exhibition or if Höch's collage influenced him. *Credit*: Bildarchiv Preussischer Kulturbesitz, Berlin.

Republic's Minister of Education, Konrad Haenisch (1876-1925).[55] Indeed, Haenisch sent Einstein a letter of support in September 1920, which was published in the press,[56] even though his aides had warned him against confusing politics and science even more than they already were.[57] The debates on the possible reforms of the DPG were fueled, on the one hand, by a genuine sense of resentment on the part of physicists living in the Reich's provinces, who felt that they were being subordinated to those living in Berlin, and, on the other hand, by their divisions along political lines, as represented by the two extremes of Lenard and Einstein.

Einstein took a rather long time--some ten days--before he started to reply to the many people who had written letters of support to him, which suggests that he needed a moment to recuperate from the strain and to reflect on his responses. Haenisch was one of the first to receive an answer, on September 8, 1920: Einstein had decided not to leave Berlin.[58] To Sommerfeld Einstein had admitted two days earlier that he perhaps should not have written his *Antwort*, but he had wanted to avoid the impression that his silence would be taken as assent.[59] Einstein also expressed his regret about his *Antwort* to Ehrenfest, but he again emphasized that he felt a need to inform the public of his opinion about the charges that had been repeatedly leveled against him.[60] By now, however, Einstein believed that "the anti-relativity company has pretty much broken down," and he jokingly remarked that he was not going to "flee the flag, as Sommerfeld puts it." In fact, the only other anti-relativity lecture that was delivered at the Berlin Philharmonic was one by the engineer Ludwig Glaser (1889-?) on September 2; the other speaker for that evening, the philosopher Oskar Kraus (1872-1942), had canceled his lecture. It thus gradually became clear to Einstein and his friends that the membership of the Working Society was not so large after all. Indeed, no other lectures would be given. Weyland had asked Willem Julius (1860-1925), a Dutch friend of Einstein's, to join his campaign (earlier in 1920 Julius had presented a paper to the Amsterdam Academy in which he had argued that the gravitational redshift had not been observed), but Julius had declined to do so.[61] Einstein now began to lighten up. As he told Julius:

> My opponents had the splendid idea to engage my dear friends for their business. You can imagine how I had to laugh when I read this...! I came to see the funny side of this whole affair a long time ago and no longer take any of it seriously.[62]

**The Bad Nauheim Meeting**
Einstein's cheerful mood, however, soon darkened again. On September 20, 1920, Friedrich von Müller, as chairman, opened the 86th meeting of the German Society of Scientists and Physicians in Bad Nauheim, and although he did speak out against the "popular assemblies with demagogic slogans" that had been directed against Einstein in Berlin, he also gave vent to his strongly nationalistic sentiments, which were warmly received by the audience. Nor was von Müller the only speaker who voiced his conservative convictions.[63] The extended debates over the restructuring of the DPG, during which Wien gained the upper hand,[64] added to a tense and reactionary atmosphere. All of which surely contributed to Einstein's sense of uneasiness.

In his *Antwort*, Einstein had challenged his opponents to a debate in Bad Nauheim.[65] In fact, months before the antirelativity campaign had gotten under way at the Berlin Philharmonic, Einstein had proposed to hold a general discussion on relativity at the Bad Nauheim meeting in lieu of giving a lecture there.[66] Now, however, following the events in Berlin, everyone-- including Germany's eager journalists--expected a sensational *Einsteindebatte*.[†] Some five-to-six-hundred people streamed toward Bad Nauheim's Bath House Number 8 (figure 5) to attend the debate.[67] Seating was limited: One person who attended reported that at 8:15 A.M. only one door opened:

---

[†] The term *Einsteindebatte* was used both by the press and by participants at the meeting; see "Die Einstein-Debatte auf dem Naturforschertag," *Berliner Tageblatt* (September 24, 1920, Morning Edition), and Körner, "86. Versammlung" (ref. 64), p. 80.

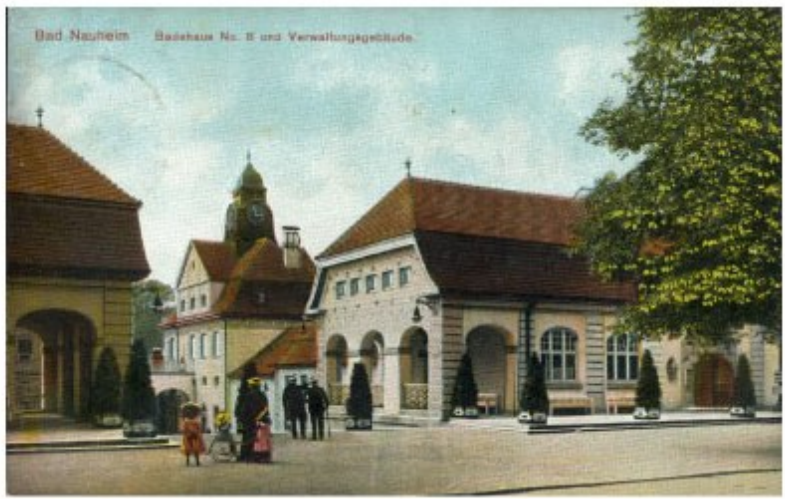

Figure 5: Bath house number 8 in Bad Nauheim, the site of the session on relativity and the Einstein-Lenard debate at the 86th meeting of the Society of German Scientists and Physicians (*Gesellschaft Deutscher Naturforscher und Ärzte*) on September 23, 1920, as pictured on a postcard of 1911.

> … and it was really quite narrow, and made even narrower because to its right and left stood a mathematician and a physicist--famous people, and not of the most slender build--like angels with broadswords standing before the Einstein paradise. Entrance to a "business meeting" was only granted by this mathematical-physical filter to members of the German Mathematical Society and the German Physical Society ... until those waiting outside could come in at 9 o'clock. They quickly filled up the chairs, stood along the walls and filled the gallery--and waited for the scholarly dispute.[68]

The audience had to wait nearly four more hours before the debate began: It was preceded by lectures by the mathematician Hermann Weyl (1885-1935) and the physicists Gustav Mie (1868-1857), Max von Laue, and Leonard Grebe (1883-1967), which were followed by a discussion,[69] until finally the floor was opened for a general debate--which lasted only fifteen minutes.

Thus began the famous "duel" or "cockfight" between Einstein and Lenard.[70] Unfortunately, no known complete account of their debate exists. The *Physikalische Zeitschrift* published a careful rendering of the arguments they exchanged--except for a hiatus halfway through it.[‡] Hermann Weyl published the other official account of the meeting for its cosponsor, the German Mathematical Society (*Deutsche Mathematikervereinigung*), but his account of the debate is colored by his reworking of its arguments and therefore is not a literal transcription of it.[71] Other participants offered their recollections later in life, but these are inconsistent and differ on matters of fact as reported in contemporary newspapers.[72]

In any case, the various accounts make clear that the most important point raised by Lenard (figure 6) was his contention that Einstein needed fictitious gravitational fields to insure the validity of his principle of equivalence.[§] Lenard exemplified this issue with the case of a "braking train": If, according to the principle of equivalence, the slowing down of a moving train is equivalent to the effect of some gravitational field, then what masses had generated this field? Einstein replied that the gravitational field responsible for slowing the train down is a perfectly valid solution of the field equations for a certain configuration of distant masses, that is, for an appropriate set of boundary conditions. Lenard responded that this just explained away a valid concern by a purely formal argument. He believed that relativity violated the intuitive understanding of physicists, principally because it dismissed the ether. Einstein then pointed out that the intuitions of physicists had changed over time.

---

[‡] The editor of *Die Naturwissenschaften*, Arnold Berliner, blamed the publisher of the *Physikalische Zeitschrift*, Hirzel, for the incomplete account of the debate; see Arnold Berliner to Einstein, January 17, 1921, Einstein Archive, Hebrew University, Jerusalem, accession no. 7-007.

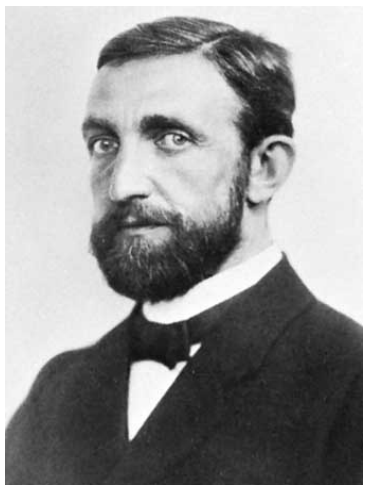

Figure 6: Phillip Lenard in 1900.

Some newspaper reports emphasized that the arguments were exchanged in an entirely objective fashion: One reporter even believed that an "exemplary ... calm" had prevailed during the debate.[73] Nonetheless, tensions had been mounting. Max Planck was firmly in the chair, but prior to the debate--because he was still not certain whether Einstein would remain in Berlin--he appeared to be quite agitated.[74] Paul Weyland also was present at the debate--but this time he kept a low profile. Einstein and his wife Elsa were strongly affected by the exchange: Elsa suffered a nervous breakdown.[75] The Viennese experimental physicist Felix Ehrenhaft (1879-1952) recalled that he had to take a highly upset Einstein out for a calming stroll in the park after the debate. Later that evening they avoided the uneasy company of their fellow physicists.[76]

Both Lenard and Einstein left the conference deeply distressed. Lenard renounced his membership in the DPG--and even denied admittance to his office at the University of Heidelberg to any of its members.[77] He now claimed that Einstein had a close connection to Moscow, and that he had tried to intimidate opponents of relativity with threatening letters.[78] Attempts at reconciliation by Planck and the Freiburg physicist Franz Himstedt (1852-1933) were to no avail.[79] Lenard's antagonism toward Einstein grew ever stronger until, in Nazi Germany, it found its place in his racist ideology of *Deutsche Physik*.[80] Einstein, on his part, deeply resented Lenard, believing that he, and also Wilhelm Wien:

---

[§] Lenard's two other principal objections centered around the abolition of the ether and a presumed contradiction in the case of a rotating frame of reference where speeds greater than the speed of light would result; see Weyl, "Relativitätstheorie" (ref. 72), pp. 59-60; "Vorträge und Diskussionen" (ref. 70), pp. 666-668; 351-356.

had only been making trouble out of a love for trouble.... I will not allow myself to be made so upset again as I was in Nauheim. I absolutely can not understand that because of bad company I could lose myself in such deep humorlessness.[81]

**Conclusions**

After these events in Berlin and Bad Nauheim, Einstein and those around him developed a heightened sensitivity to the unwelcome consequences of publicity. When Alexander Moszkowski--who had hailed relativity in the *Berliner Tageblatt* as a new truth that surpassed those of Galileo, Newton, and Kant--was about to publish a book based on his conversations with Einstein,[82] Max Born (1882-1970) and his wife Hedwig (1891-1972), and Einstein himself, tried unsuccessfully to stop its publication. They feared that it would spark an entirely new campaign against Einstein: The book at one stroke would confirm Weyland's charge of shameless self-promotion.[83]

Yet, despite these public pressures, Einstein remained in Germany. Despite the publicity to which the newspapers had subjected him; despite the personal and at times racist attacks he had endured; despite the conservative and nationalistic convictions and politics of German academics who surrounded him; despite the rampant inflation after the war; despite the poverty and social instability--Einstein stayed in Berlin. He could have accepted an offer to return to Switzerland,[84] or he could have encouraged Ehrenfest to indeed secure a full professorship for him in Leiden, which he greatly enjoyed visiting.[85] Yet, Einstein chose to remain in Berlin--not, of course, for any patriotic reasons, although some people attempted to appeal to them. He realized, however, that if he were to depart, Germany would be disgraced,[86] and his Berlin colleagues would be embarrassed or even feel that they had been betrayed. He felt a responsibility--an uncomfortable one--to avoid such consequences.[87]

Einstein also knew that he had to support his first wife Mileva Marić (1875-1948) and his children, who were living in expensive Switzerland. He feared that even if this would make his financial situation untenable, he would not be able to muster enough courage to leave Berlin and accept a position elsewhere. He expressed his frustration to his old friend Marcel Grossmann (1878-1936):

[P]eople here will make desperate attempts to keep me. To a lesser extent because they want to have me for personal reasons and because of my brain ooze, to a larger extent because I have become a fetish due to the newspaper clamor. I play a part similar to that of the bones of the saint, which one must have in the convent's church. My departure would be felt as a lost battle.[88]

Einstein also felt tied to Berlin by his strong personal friendships and the genuine professional appreciation he experienced there. "Here for the first time I lost the feeling of painful loneliness from which I always suffered in Bern and Zurich."[89] Einstein's friendships in Berlin supported him in his sense of duty, and even offered the calmness to allow him to put the machinations of Weyland and Lenard in perspective:

The problems and attacks that here have come my way are not so bad as they may appear abroad. It would also be an injurious act when in this time of distress and humiliation I would turn my back on Germany, given the great kindness that I have constantly experienced from the side of my German colleagues and authorities. That I am Swiss and generally have an international outlook, does not change this personal relation. I therefore consider it my duty to endure in my position until outside circumstances render it practically impossible.[90]

## Acknowledgments

I presented a first version of this paper at the Seventh International Conference on the History of General Relativity, Tenerife, Spain, on March 11, 2005, and I thank the organizers, the Fundacion Orotava de Historia de la Sciencia and the Max Planck Institute for the History of Science, for their kind hospitality. I am grateful to Diana Kormos Buchwald for drawing my attention to Hannah Höch's collage and to her and the other members of the editorial team of Volume 10 of the *Collected Papers of Albert Einstein*, Tilman Sauer, Jószef Illy, Ze'ev Rosenkranz, Virginia Iris Holmes, A.J. Kox, and Daniel Kennefick, for a most rewarding collaboration with them. I also am grateful to The Netherlands Organization for Scientific Research (NWO) for its generous support through a VENI grant. Finally, I thank Roger H. Stuewer for his careful and thoughtful editorial work on my article.